\documentstyle[prl,aps,epsf,rotate,multicol]{revtex}

\include{epsf}

\title{The Random Fuse Network as a Dipolar Magnet}

\author{Marc Barthelemy$^1$, Rava da Silveira$^2$, and Henri
Orland$^3$}

\address{
$^1$ CEA, Service de Physique de la Mati\`ere Condens\'ee, BP12, 91680
Bruy\`eres-le-Ch\^atel, France.\\
$^2$ Department of Physics, Harvard University, Cambridge, 
Massachusets 02138, USA.\\
$^3$ CEA, Service de Physique Th\'eorique, 91191 Gif-sur-Yvette,
Cedex, France.}

\begin{document}
\draft

\maketitle

\begin{abstract}

We introduce an approximate mapping between the random fuse network
(RFN) and a random field dipolar Ising model (RFDIM). The state of the
network damage is associated with a metastable spin configuration.  A
mean-field treatment, numerical solutions, and heuristic arguments
support the broad validity of the approximation and yield a generic
phase diagram. At low disorder, the growth of a single unstable
`crack' leads to an abrupt global failure. Beyond a critical disorder,
the conducting network sustains significant damage before the
coalescence of cracks results in global failure.

\end{abstract}



Heterogeneities are present in real materials, in
a wide range of magnitudes and scales, and generate fluctuations in
local toughness. The fracture mechanics then results from an interplay
between disorder and stress
fluctuations\cite{Roux90,Chakrabarti97}. In attempting to elucidate
the role of the disorder, a number of classifications and
characterizations of rupture have been proposed
\cite{Hansen91,Selinger91,Buchel97,Zapperi97,Kahng88,Moukarzel94,Andersen97,Sahimi86}. In
particular, the behavior of the damage as the applied force approaches
its critical value (global rupture)---partial versus no damage,
macroscopic abrupt versus continuously increasing damage---allowed to
distinguish various types of fracture. Most studies were carried out
either numerically or in the framework of extreme statistics (see
Refs.~\cite{Roux90,Chakrabarti97} and references therein), while
few satisfactory analytical approaches exist. In parallel, a number of
authors\cite{deArcangelis85,Takayasu85} introduced the so-called
random fuse network (RFN), less involved than fracture models because
of its scalar nature but still capturing the main aspects of rupture
mechanics\cite{Gilabert87}. Thus the RFN often is seen as a toy model
for the investigation of mechanical rupture. The mechanical bonds are
replaced by fuses burning if the current (or equivalently the
electrical field) exceeds a random threshold. As the total applied
current flowing through the network is increased, more and more fuses
burn, thereby decreasing the effective conductivity of the network
until a global failure at which no current can go through anymore. The
similarity of the RFN with a driven random field Ising model at zero
temperature\cite{Dahmen96} has been recently noted\cite{Zapperi97},
but no systematic correspondence has been established.

In this letter, we present a mapping, valid at least at the initial
stages of failure, between the RFN and a driven random field Ising
model with {\it dipolar} coupling. This mapping associates the
configuration of the burnt fuses at a given value of the applied
electric field with a zero temperature metastable state of the random
field dipolar Ising model (RFDIM). The mapping between fracture and a
classical model of statistical mechanics, is supported by mean-field
and heuristic arguments and by a preliminary numerical study, from
which a generic phase diagram with a disorder induced transition
emerges.


The RFN consists of a $d$-dimensional lattice of fuses, each with a
conductivity $\sigma_i(\bf{x})=\sigma_0$ initially, where ${\bf x}=a
{\bf m}$, ${\bf m}\in{\bf Z}^d$ and $a$ is the lattice spacing (we
will assume the continuum limit $a\rightarrow 0$), and $i=1,\ldots,d$
corresponds to the orientation of the fuse. If the electric field
along $i$ at ${\bf x}$, $E_i({\bf x})$, exceeds a random threshold
${\cal E}_i({\bf x})$, the fuse burns and $\sigma_i({\bf x})=0$. We
consider a scenario in which the electric field in the direction $i=1$
is slowly increased from zero and $E_i=0$ for $i\not= 1$ at all times.

Assuming locality and analyticity, we can write a general 
equation for the time evolution of the conductivity field as
\begin{equation}
\eta \partial_t
\sigma_i({\bf x},t)=
f_i(\{\sigma_j({\bf x},t)\},
\{\partial_j\sigma_k({\bf x},t)\},\ldots)
\theta(-\Delta_i({\bf x},t)),
\end{equation}
where $\Delta_i={\cal E}_i({\bf x})^2 -E_i({\bf x},t)^2$ and the
Heaviside function ensures that the conductivity of the fuse remains
$\sigma_0$ as long as the electric field it sustains does not exceed
its threshold. The simplest choice for the function $f_i$ within the
constraints of positivity $(\sigma_i({\bf x},t)\geq 0)$ and
monotonicity $(\partial_t\sigma_i({\bf x},t)\leq 0)$, $f_i=-~ {\rm
constant} ~\times \sigma_i({\bf x},t)$, yields
\begin{equation}
\eta\partial_t\sigma_i({\bf x},t)=-
\sigma_i({\bf x},t)
\theta(-\Delta_i({\bf x},t))
\label{eqnum1}
\end{equation}
after absorption of the constant in a redefinition 
of $\eta$. Equation (\ref{eqnum1}) incorporates two time scales:
the relaxation time $\eta$ and the characteristic time
$\tau\equiv E_1(dE_1/dt)^{-1}$. We focus on the limit
$\eta/\tau\rightarrow 0$, in which the conductivity
relaxes instantaneously from $\sigma_0$ to 0
as the local field crosses the local threshold. To
capture the sequential dynamics associated with the 
$\eta\not= 0$ retardation effect, we 
discretize \cite{fish} Eq.~(\ref{eqnum1}) in time as
\begin{equation}
\sigma_i({\bf x},t+\delta t)=
\sigma_i({\bf x},t)\theta(\Delta_i
({\bf x},t)).
\label{eqnum2}
\end{equation}
Formally, this corresponds to setting $\eta=\delta t$ and discretizing
Eq.~(\ref{eqnum1}) in the usual fashion. As expected,
Eq.~(\ref{eqnum2}) prescribes an irreversible dynamics such that
$\sigma_i({\bf x})=0$ at time $t$ implies $\sigma_i({\bf x})=0$ for
all later times.

For a given configuration of the conductivities, 
Kirchoff's laws may be recast \cite{Hori77}
into the integral equation
\begin{equation}
\label{Eequation}
E_i({\bf x}) =E_0\delta_{i1}+
\int d^dy
\sum_j G_{ij} 
({\bf x}-{\bf y})
\delta\sigma_j({\bf y}) E_j({\bf y}),
\end{equation}
where $E_0$ is the magnitude of the external field applied along the
first axis, $\delta\sigma_i({\bf x})= \sigma_i({\bf x})-\sigma_0$,
$G_{ij}({\bf x})=- \partial_i\partial_j g({\bf x})$ the dipolar
tensor [17a], and the Green's function $g$ is defined by
$\sigma_0\nabla^2 g({\bf x})=-\delta({\bf x})$. To lowest non-trivial
order, Kirchoff's laws become
\begin{equation}
\label{Eapprox}
E_i({\bf x})=E_0\left[\delta_{i1}+\int d^dy G_{i1}
({\bf x}-{\bf y})\delta\sigma_1({\bf y})\right].
\end{equation}
As we shall see, this simple approximation captures much of the
physics of network damage. First, due to the presence of the dipolar
kernel $G_{ij}$, Eq.~(\ref{Eapprox}) reproduces the expected shielding
of the current `upstream' and `downstream' to a burnt region, as well
as the lateral amplification \cite{Gilabert87,Kachanov94}. Second, the
$k$th term of the expansion contains a product
$\prod_{i=1,k}\delta\sigma_1(y_i)$ which vanishes with high
probability, especially at the early stages of the damage. In
addition, this approximation satisfies the necessary property that the
field at the tip of a crack increases with crack growth. For one
isolated burnt fuse, the field on the next parallel link in the
direction perpendicular to the applied field is enhanced by a factor
$\alpha > 1$. The approximation (\ref{Eapprox}) gives
$\alpha=1-G_{11}(0,1)\simeq 2.14$ on the $2d$ square lattice (and
$\alpha=2$ in the continuum limit [17b]); the exact result on the
square lattice is $\alpha=4/\pi$. When $n$ such parallel fuses are
broken, the field at the tip of the crack is known to scale as
$n^{1/2(d-1)}$\cite{Duxbury87}, while in the approximation of
Eq.~(\ref{Eapprox})
\begin{equation}
E_{\rm tip}=E_1(0,\rho=n+1)=E_0(1+\sum_{1<\rho<n+1} 1/\rho^d)
\end{equation}
(where $\rho^2=x_2^2+\dots+x_d^2$ is the transverse square distance)
converges to a finite value as $n$ increases, corresponding formally
to the $d=\infty$ case. For uniform disorder distributed according to
$p({\cal E})=\theta({\cal E}^2-{\cal E}_m^2)\theta({\cal
E}_m^2+w-{\cal E}^2)/w$ with $w<2$ or a binary disorder $E_{\rm tip}$
can be larger than the largest threshold and we expect the same
results as if $E_{\rm tip}\sim n^{1/2(d-1)}$. For a disorder
distribution allowing with a high probability very large values of
thresholds (such as distributions with fat tails), the scaling of
$E_{\rm tip}$ with the size of the crack becomes relevant and we
cannot use the approximation (\ref{Eapprox}). Here, we present results
for the uniform disorder while similar results are obtained for binary
disorder.

Squaring Eq.~(\ref{Eapprox}) we obtain, to lowest nontrivial order, the
distance $\Delta_i$ from the local threshold in terms of the damage
$\delta\sigma_i$, as
\begin{eqnarray}
\label{approxa}
\Delta_1({\bf x},t)&\simeq&{\cal E}_1({\bf x})^2-E_0^2 -2E_0^2\int d^dy\
G_{11}({\bf x}-{\bf y})\delta\sigma_1({\bf y},t)\\ 
\Delta_{j}({\bf x},t)&\simeq&{\cal
E}_j({\bf x})^2 -E_0^2\left[\int d^dy\ G_{j1}({\bf x}-{\bf
y})\delta\sigma_1({\bf y},t)\right]^2,
\label{approxb}
\end{eqnarray}
(with $j\neq 1$) which, along with the update rule of Eq.~(\ref{eqnum2})
completely determines the state of the network under an applied field
$E_0$.

In terms of the `Ising spins' defined by $s({\bf
x},t)=2\delta\sigma_1({\bf x},t)/\sigma_0+1$, Eq.~(\ref{eqnum2})
simply stipulates that each spin $s({\bf x},t)$ will lie along a local
field $\Delta_1({\bf x},t)$. Eq.~(\ref{approxa}) shows that this local
field $\Delta_1$ results from a uniform applied component, a random
component, and a spin-spin interaction mediated by the component
$G_{11}$ of the dipolar kernel. Therefore, as the applied field is
switched from $0$ to $E_0$, spins flip in avalanches until they reach
the state $s({\bf x})={\rm sgn}[\Delta_1({\bf x},t=\infty)]$ which
minimizes the Hamiltonian
\begin{eqnarray}
\label{hamiltonien}
{\cal H} = - \frac{1}{2}\int d^dx d^dy
\/s({\bf x})J({\bf x}-{\bf y})s({\bf y})-\int d^dx\/h({\bf x})s({\bf x}),
\end{eqnarray}
where $J({\bf x})=-E_0^2\sigma_0G_{11}({\bf x})$, $h=H+{\cal E}_1({\bf
x})^2$, and $H=-E_0^2[1-\sigma_0{\tilde G}_{11}({\bf
0})]=-E_0^2(1+1/d)$. As the driving field $|H|$ is increased, more and
more spins flip downward, and the state of the RFN follows the
metastable (nonequilibrium) spin state of ${\cal H}$, connected by its
history to the initial condition $s({\bf x})=+1$ for all ${\bf x}$.

While similarities between the RFN and the random field Ising model
(RFIM) have been pointed out in the past, here a precise mapping is
elucidated. We stress that, at odds with previous studies on the
RFIM\cite{Young97} in which the spin-spin coupling is taken to be
short-ranged, constant, and isotropic, the coupling $J({\bf x})$ is
rather peculiar as its magnitude is proportional to the applied field
$H$ and its modulation is dipolar. Even in the absence of disorder,
the dipolar interaction in Ising spin models\cite{DeBell00}
leads to a rich behavior \cite{Nattermann88,Magni00} not yet fully
understood. In particular, a striped phase was identified
\cite{Garel82,MacIsaac95} with alternating up and down spins. A
similar scenario occurs in the RFN at very low (narrow) disorder. All
the fuses remain intact until $E_0$ reaches the lowest threshold and
the corresponding fuse burns. The dipolar coupling then relieves its
longitudinal neighbors while further stressing its transverse
neighbors, who burn in turn, and a stripe of burnt fuses develops,
eventually spanning the whole system (the network then stops
conducting and the formation of additional stripes is consequently
prohibited). At high (wide) disorder, a resilient fuse in the passage
of a burnt stripe will stop the growth of the crack, and new cracks
will nucleate elsewhere. This suggest a transition, upon increase of
the disorder\cite{Kahng88,Andersen97,Vives01}, from an abrupt regime
characterized by the growth of a single macroscopic unstable crack to
a continuous regime in which the system is significantly damaged
before its global failure.

For a qualitative understanding of the role of disorder, we first
revert to a simpler version of the problem in which the dipolar kernel
is replaced by a uniform infinite-range coupling $J({\bf x})=E_0^2
J/N$, where $J>0$ measures the strength of the interaction and $N$ is
the total number of spins (equivalently, fuses). This `democratic'
model is mean-field like in that the evolution of the local field
(Eq.~(\ref{approxa})) acting on a spin $s({\bf x})$ may be described
in terms of a single degree of freedom, the fraction $n_-$ of down
spins (burnt fuses),
\begin{equation}
\Delta_1({\bf x})={\cal E}_1({\bf x})^2-E_0^2-2JE_0^2n_-,
\end{equation}
and depends on the position only through the local random
threshold. Upon increase of the applied field a number of spins flip,
thus increasing $n_-$ which in turn causes more flips, and so on until
this avalanching terminates at $n_-(E_0)$, the metastable fraction of
downward spins at an applied field $E_0$. We follow the graphical
scheme of Ref. \cite{silv} to obtain the resulting phase diagram. For
the uniform disorder of width $w$, clearly $n_-=0$ as long as
$E_0^2<{\cal E}_m^2$, the lowest threshold. If $w<2J{\cal E}_m^2$,
$n_-$ then jumps to 1 abruptly; if $w>2J{\cal E}_m^2$, on the other
hand, $n_-=(E_0^2-{\cal E}_m^2)/(w-2JE_0^2)$ increases continuously
with $E_0$ up to $n_-=1$ at $E_0^2=({\cal E}_m^2+w)/(1+2J)$. As
expected, a critical width $w_0\equiv 2J{\cal E}_m^2$ of the disorder
separates two regimes: at low disorder, a huge avalanche is leading to
global failures, while for large disorder, tiny avalanches can be
stable, until cracks coalesce.  We note the broad validity of this
picture, at least in the mean field. In contrast with the `democratic
fiber bundle model' \cite{peir,dan} for which it was shown \cite{silv}
that abrupt failures are but an artifact of a large discontinuity in
the threshold distribution, here the phase diagram extends to
distributions that are not continuous, e.g., to any uniform
distribution on a support $[{\cal E}_m^2,{\cal E}_m^2+w]$. These
mean-field results are also recovered from the homogeneous saddle
point of the partition function 
\begin{equation}
Z=\sum_{\{s({\bf x})=\pm
1\}}e^{-\beta{\cal H}}
\end{equation}
This saddle point corresponds to the stability condition for a
microcrack: if the electric field at the tip of the crack is larger
than the largest threshold, the crack is unstable. This implies the
existence of the critical width $w_0$ found above. It is quite
remarkable that this result---which was already found by probabilistic
methods\cite{Kahng88}---appears here naturally as the uniform saddle
point or mean field.

Because of the form of the coupling in this simplified democratic
model, we can carry out a similar investigation without recourse to
the linear approximation of Eq.~(\ref{Eapprox}), but using
Eq.~(\ref{Eequation}) directly. Assuming $J<1/2$ (without which the
problem is ill-defined), we find not only a similar phase diagram but
also, surprisingly, the same value for the critical disorder
$w_0=2J{\cal E}_m^2$. At least in this mean field version of the
original problem, the linear approximation is legitimate. This also
suggests that when Eq.~(\ref{Eapprox}) is valid, the early stages of
the damage dominate the whole breakdown process.

Evidently, a mean field phase diagram may be significantly modified by
fluctuations governed by a dipolar coupling. On the one hand the
dipolar kernel decays as $1/r^d$, on the other hand its angular
dependence invests the model with a so-called nonmonotonicity: burning
a fuse relieves the current on a fraction of neighboring
fuses. Monotonicity is a simplifying feature of many driven systems,
and its lack clearly introduces complications and a possibly richer
phenomenology\cite{Vives01}. Furthermore, while a mean field treatment
predicts the evolution of damage, it fails to capture the random
fluctuations of the breakdown field $E_b$ as well as finite-size
effects. These two features are however present in the approximation
(\ref{Eapprox}). In particular, it is easily shown that the maximum of
the electric field scales as $\ln L$, as a consequence of the long
range of the dipolar tensor. Investigation of finite size effects in
the framework of this approximation present a complementary route to
most of the approaches to finite size effects, which rely on extreme
statistics\cite{Duxbury86} and neglect correlations.


For a better grasp of the full problem we obtained numerical solutions
of the RFN (using Eqs.~(\ref{eqnum2},\ref{Eequation})) and the RFDIM (using
the approximation Eq.~(\ref{approxa})). We found (Fig.~1) that the
RFDIM reproduces quite well the behavior of the RFN. In both cases, we
observe a critical value $w_0$ of the disorder beyond which there is a
large window of damage that preceeds the global failure.


Before concluding, we propose a first attempt to describe the
stability of cracks in the spirit of the Imry-Ma
argument\cite{Imry75,Nattermann88}. In three dimensions, the creation
of a spherical crack of radius $R$ modifies the total energy
(cf. Eq.~(\ref{hamiltonien})) by
\begin{equation}
\Delta {\cal H}\approx 2\overline{h}R^3-wR^{3/2}+gR^3
\end{equation}
where the last term comes from the dipolar interaction ($g>0$) and
merely renormalizes the average of the random
field\cite{Nattermann88}. If the applied field $E_0$ is small, the
average $\overline{h}$ is positive and the formation of a large crack
is prohibited. In this case, the typical size of a crack will be
$R_0\approx [w/(2\overline{h}+g)]^{2/3}$. On the other hand, if
$2\overline{h}+g<0$ it is favorable for a crack to grow
indefinitely. A more precise argument based on an oblate
(`penny-shaped') crack perpendicular to the applied field leaves the
above conclusions unchanged. As $w$ is increased, the typical size
$R_0$ of damaged regions becomes larger, up to a critical value $w_0$
at which the latter percolate. This occurs when $R_0$ becomes
comparable to the typical distance between nucleation sites,
$a[w/(H-{\cal E}_m^2)]^{1/3}$ for an uniform disorder. As long as
$w<w_0$, breakdown results from a single crack that spans the system
when $2\overline{h}+g$ becomes negative upon increase of the applied
field $H$. When $w>w_0$, localized cracks grow and percolate before
$2\overline{h}+g$ changes sign, which allows significant damage while
the network is still conducting. This heuristic argument also
substantifies the broad validity of the approximation of
Eq.~(\ref{Eapprox}) {\it a priori} only valid at the initial stages of
the damage: the breakdown is controlled either by an instability or by
the coalescence of (small) cracks.


In summary, we have presented an approximate mapping of the RFN to the
RFDIM, and argued for its broad validity. The state of the RFN---in
particular the extent of damage (number of burn fuses)---is mapped
into a metastable spin state. From a mean-field investigation,
numerical solutions, and heuristic arguments, a generic picture
emerges, characterized by a critical value of the disorder. At low
disorder, a macrosocopic crack grows until it spans the system. At
large disorder, threshold fluctuations stabilize (micro)cracks and a
significant precursor damage develops before the latter coalesce into
a percolating structure. Our investigation were carried out for a
bounded disorder. By contrast, a slowly decaying threshold
distribution may modify the character of the transition or suppress it
altogether, an interesting question for future study.


\acknowledgments
We thank T.~Garel, M.~Kardar, Y.-P.~Pellegrini, D.~Rittel, S.~Zapperi
for interesting discussions. RdS is supported by a Young Scientist
Grant from the Swiss National Science Foundation, the NSF through
Grant No.  DMR-98-05833, and the Harvard University Society of
Fellows.



\newpage

\begin{figure}[t]
\narrowtext
\centerline{
\epsfysize=0.6\columnwidth{\epsfbox{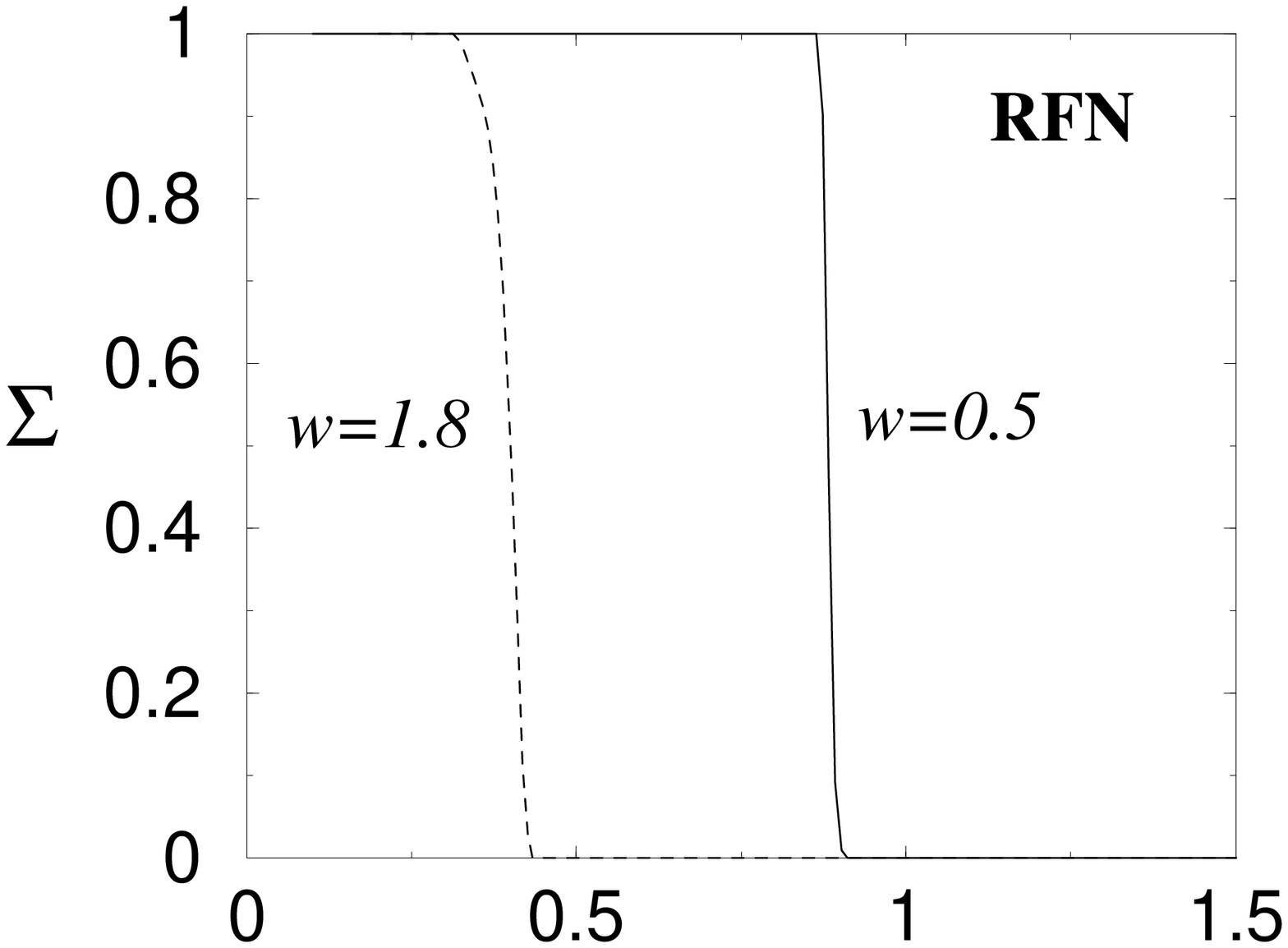}}}
\vspace*{0.2cm}
\centerline{
\epsfysize=0.65\columnwidth{\epsfbox{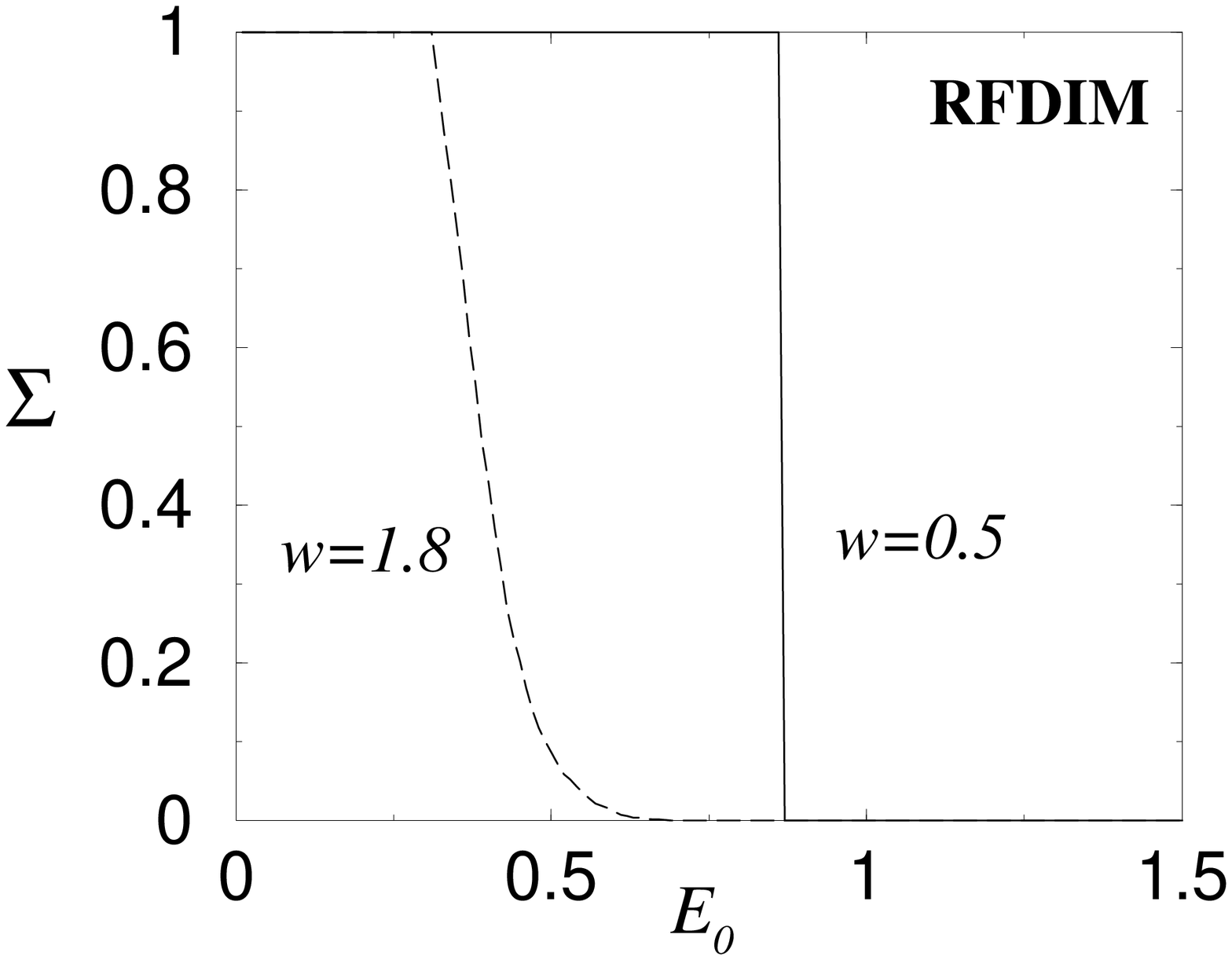}}}
\vspace*{0.2cm}
\caption{Effective conductivity $\Sigma$ versus $E_0$ for uniform
disorder for two values of the disorder width $w=0.5<w_0$ and
$w=1.8>w_0$ (for a system size $50\times 50$ and averaged over $100$
configurations). (a) Calculation for the RFN by solving
Eqs.~(3,4). (b) Calculation for the RFDIM obtained from the numerical
solution of Eqs.~(6a,b). In both cases (a) and (b), for $w<w_0$, the
breakdown is abrupt and there is no fluctuation in the breakdown field
(for (a), it is not fully abrupt, due to finite-size effects). For
$w>w_0$, the breakdown field fluctuates and there is some damage
before complete breakdown of the system.}
\label{figure1}
\end{figure}



\end{document}